\documentclass[12pt]{iopart}
  
\usepackage[utf8]{inputenc}
\usepackage{graphicx}
\usepackage{blindtext, graphicx}   

 \usepackage{amstext}
 
\usepackage{lipsum}   
\usepackage{float}   

\usepackage{xcolor}
\usepackage{soul}
\usepackage{ulem}
\usepackage{cancel}

\usepackage[mathlines]{lineno}

\begin{document}
\title[Random-search efficiency]{ 
Random-search efficiency in a bounded interval with spatially heterogeneous diffusion coefficient
}

\author{L. Menon Jr.$^{1}$, M. A. F. dos Santos$^{1}$  and C. Anteneodo$^{1,2}$}
\address{$^{1}$ Department of Physics, PUC-Rio, Rua Marqu\^es de S\~ao Vicente 225, 22451-900, Rio de Janeiro, RJ, Brazil}
\address{$^{2}$ Institute of Science and Technology for Complex Systems, INCT-CS Brazil}

\begin{abstract}
We consider random walkers searching for a target in a bounded one-dimensional heterogeneous environment, in the interval $[0,L]$, where diffusion is described by a space-dependent diffusion coefficient   $D(x)$. 
Boundary conditions are absorbing at the position of the target (set at $x=0$) and reflecting at the border $x=L$.  We calculate and compare the estimates of efficiency   $\varepsilon_1=\langle 1/ t\rangle$ and $\varepsilon_2=1/\langle t \rangle$. 
For the Stratonovich framework of the multiplicative random process, both measures are analytically calculated  for arbitrary $D(x)$. 
For other  interpretations of the stochastic integrals (e.g., It\^o and anti-It\^o), we get general results for $\varepsilon_2$, while $\varepsilon_1$ is obtained for particular forms of $D(x)$.        
The impact of the diffusivity profile on these measures of efficiency is discussed.  
Symmetries and peculiar properties arise when the search starts at  the border ($x_0=L$), in particular, heterogeneity spoils the efficiency of the search within the Stratonovich framework, while for other interpretations the searcher can perform better in certain heterogeneous diffusivity profiles.
 
\end{abstract}


\section{Introduction} \label{sec:intro}

The study of random searches is valuable in several areas of science and technology~\cite{chupeau2015}, 
offering effective strategies for exploring the accessible space and finding targets. 
Random searches can be found from the molecular scale, as in problems of protein-DNA binding~\cite{mirny2009protein,chen2019target,bhattacherjee2014search}, to ecological scales,  where random searches are crucial in foraging for resources  or viable  habitats, necessary for survival and reproduction~\cite{LuzNature1999,o1990search,bartumeus2005animal,viswanathan2011physics,MartinezGarcia2013,Redner2022}. 
We can also find technological applications, as in robotic design~\cite{castello2016} or in the optimization of hyperparameters in neural networks~\cite{bergstra2012}.

 We address the  situation in which a random search occurs in an heterogeneous confined environment. We assume that the search dynamics follows a Brownian (or diffusive) motion, such that a non-uniform, i.e., position-dependent, diffusion coefficient reflects the heterogeneity of the environment. 
State-dependent diffusion appears in  the description of diverse physical systems  that exhibit spatial complexity, producing anomalous diffusion, e.g., subdiffusion~\cite{zodage2023sluggish}
and turbulent diffusion~\cite{stella2022}, 
as well as ergodicity breaking~\cite{cherstvy2013anomalous}. 
It also appears in the modeling of 
 biological~\cite{english2011single,neuralPRX}, 
 socioeconomic~\cite{vieira2018threshold}, 
and ecological \cite{dos2020critical} systems, among other examples. 

Random searches in heterogeneous media have been previously studied  in  non-confined domains, exploring different diffusivity profiles 
within Stratonovich scheme ~\cite{mutothya2021first}, for general schemes of the stochastic integrals~\cite{Santos2022}, or with stochastic resetting~\cite{Ray2020}. 
However, in real systems random searches occur in bounded domains.
As examples, 
reaction-diffusion processes take place in  limited plates~\cite{rice1985diffusion}, 
animal search for food occurs in geographically delimited habitats~\cite{LuzNature1999}, 
the visual search for an object in an image is bounded by image  size~\cite{wolfe2004}.  
In such bounded domains, the homogeneous problem has been extensively studied. 
A review about random searches in homogeneous bounded media 
can be found in Ref.~\cite{benichou2014first}, after which, extensions such as 
searches under stochastic resetting~\cite{christou2015diffusion}, 
and partially-reactive targets~\cite{bressloff2022diffusion} have been addressed. 
Studies about search processes in confined heterogeneous environments, involving position-dependent diffusion coefficient, are relatively recent, and address  
particular diffusivity profiles~\cite{li2020particle,vaccario2015first} or particular interpretations of the stochastic integrals, e.g.,  within Hänggi-Klimontovich (or anti-It\^o interpretation~\cite{godec2015optimization}). 
In this work, we aim to obtain analytical results for the search problem in  bounded one-dimensional heterogeneous environments with general forms of the state-dependent diffusivity and for general interpretation of the stochastic integrals. 

For this purpose, we consider the stochastic equation
\begin{eqnarray} \label{eq:langevin1}
\dot{x} =  \sqrt{2D(x^*)}\,\eta(t), 
\end{eqnarray}
where $\eta(x)$ represents a delta-correlated Gaussian noise and the (definite-positive) function $D(x)$ denotes the state-dependent diffusion coefficient. 
Importantly, since the white noise is multiplicative, it is necessary to choose the specific instant at which $x^*$ is calculated.  
We consider $x^*=  [(2-A)\,x(t+dt)+A\,x(t)]/2$~\cite{vaccario2015first}, where $A \in[0,2]$, focusing on three special interpretations:  Itô ($A=2$)~\cite{ito1944109},  
Stratonovich ($A=1$)~\cite{stratonovich1966new}, anti-Itô  ($A=0$)~\cite{hanggi1982nonlinear,Klimontovich1990}. 
 Alternatively, Eq.~(\ref{eq:langevin1}) can be cast in the It\^o form ($x^*=x(t)$)  
\begin{eqnarray} \label{eq:langevin2}
\dot{x} = (1-A/2)D'(x)+ \sqrt{2D(x)}\,\eta(t), 
\label{eq:langevin_A}
\end{eqnarray}
where $D'=dD/dx$  and the first term is the  noise-induced drift~\cite{volpe2016effective}. 
The Fokker-Planck equation associated to Eq.~(\ref{eq:langevin_A}) is given by~\cite{PhysRevE.99.042138}
\begin{eqnarray}
\frac{\partial \ }{\partial t} p(x,t|x_0)= \frac{\partial \ }{\partial x}\left\{ D(x)^{1-\frac{A}{2}} \frac{\partial}{\partial x}  [D(x)^{\frac{A}{2}}p(x,t|x_0) ]      \right\},
\label{eq:FPEall}
\end{eqnarray}
where  $x_0\in \Omega \equiv (0,L]$ is the initial position and  the boundary at $L$ is modelled by a reflecting boundary condition. Additionally, 
 without loss of generality we assume that the target is located at $x=0$, where we set an absorbing boundary condition,
 meaning that the searcher is removed when the search ends, upon reaching the target.

It is  interesting to note that similar forms of spatial heterogeneity can  be introduced through the potential  of a field acting on the Brownian particle~\cite{Palyulin2012,Chupeau2020}. Moreover, as can be seen in Eq.~(\ref{eq:langevin_A}), non-uniform diffusivity produces a spurious drift for interpretations other than It\^o.

To analytically grasp random searches, 
it is central  to determine the first-passage time distribution (FPTD)  
\begin{eqnarray} \label{eq:wp}
\wp(t)=-\frac{d\ }{dt} Q(x_0,t)\,,
\end{eqnarray}
where  $ Q(x_0,t) = \int_{\Omega}p(x,t|x_0)dx$,   is the survival probability at time $t$. The FPTD represents the probability density function of first-arrival times when the particles hit the target for the first time, and after which are removed from the system~\cite{redner2001guide,risken1996fokker}. 
The FPTD can be used to reveal efficient strategies that minimize the characteristic time to encounter a target, or optimize other search criteria, which can be crucial in diverse contexts and  scales~\cite{redner2001guide,benichou2011intermittent,zaburdaev2015levy}. 
As a single-value measure, the mean first-passage time (MFPT),  
$ \left \langle t \right \rangle   =  \int_0^{\infty} t\, \wp(t) dt $, 
is an important quantity to be calculated, and in some cases may be enough to characterize a random search, as well as in other first-passage problems, 
when it is finite and the corresponding standard deviation relatively small. 
In the context of search processes, associated to the so called {\it cruise motion}~\cite{john1989flexible} in which the searcher keeps exploring all the points along the trajectory,  the search efficiency is defined as number of visited targets over average number of steps, which in the case of a single target can be interpreted as the inverse of the MFPT~\cite{palyulin2014bias}, 
 
\begin{eqnarray}
\varepsilon_1=  \left \langle t \right \rangle^{-1}  = \left(\int_0^{\infty} t\, \wp(t) dt \right)^{-1}
\label{eq:1/<t>}.   
\end{eqnarray}
However, this quantity vanishes in cases where the MFPT is divergent. 
Alternatively, a different measure of efficiency has been proposed~\cite{palyulin2014levy}, namely,  
\begin{eqnarray}
\varepsilon_2=   \left \langle  t^{-1} \right \rangle = \int_0^{\infty} t^{-1}\, \wp(t) dt.  
\label{eq:inverset}
\end{eqnarray}
This measure is adequate for systems where the MFPT 
diverges~\cite{palyulin2014levy}, typically when $\wp$ has long tails, while 
these tails do not contribute to the first-order negative moment, which dismisses trajectories that take very long times to reach the target, and preferentially weights the contribution of short arrival times.  
Therefore, it is useful to consider additionally the so-called reliability, which is the complement of the survival probability, $1-Q(x_0,t\to\infty)$, at asymptotically long times, providing the fraction of particles that reached the target.  
Eq.~(\ref{eq:inverset})   has been used  to characterize the performance of L\'evy searches, facing multiple  targets~\cite{palyulin2017comparison}, under external bias~\cite{palyulin2014bias}, 
comb structures~\cite{Sandev2019} and asymmetric L\'evy flights~\cite{padash2022asymmetric}.
Let us note that other negative moments might also be considered. 

Our focus is to obtain analytically and compare the  measures of search performance defined in Eqs. (\ref{eq:1/<t>}) and (\ref{eq:inverset}),
for general types of bounded heterogeneous environments, characterized by different forms of $D(x)$ and interpretations of the heterogeneous diffusion process (HDP).    
In particular, within Stratonovich framework,  we will be able to find general results for arbitrary $D(x)$.

The remaining of the paper is organized as follows. 
 In Sec.~\ref{sec:Stratonovich},  under the Stratonovich interpretation,  we manage to calculate both  $1/\langle t \rangle$ and $\langle 1/ t \rangle$, for arbitrary profiles $D(x)$. 
In Sec.~\ref{sec:all}, we consider arbitrary interpretations ($A \in [0,2]$). In this case, for general $D(x)$, we  obtain and discuss a closed expression for the MFPT, 
hence for $\varepsilon_1$,
while for the efficiency given by the first negative moment
$\varepsilon_2$, analytical expressions are obtained for  particular shapes of $D(x)$. 
From the analytical results, the impact of the diffusivity profile and interpretation of the stochastic integrals on search performance is discussed.
Final remarks are presented in Sec.~\ref{sec:final}.

\section{Search efficiency under Stratonovich interpretation }
\label{sec:Stratonovich}

The backward FP equation associated to Eq.~(\ref{eq:FPEall}),  
under Stratonovich interpretation ($A=1$), is given by~\cite{risken1996fokker}  
 
\begin{eqnarray}
\frac{\partial \ }{\partial t} Q(x_0,t)  =   \sqrt{D(x_0)} \frac{\partial \ }{\partial x_0} \left\{ \sqrt{D(x_0)} \frac{\partial \ }{\partial x_0}  Q(x_0,t) \right\},  
\label{eq:backwardA1}
\end{eqnarray} 
where $Q(x_0,t)=  \int_{\Omega}p(x,t|x_0)dx$ is the survival probability. 
The position of the target, located at $x=0$, can be represented by an absorbing boundary condition, while the confinement of the accessible domain is given by a reflecting boundary at $x=L$, namely, 
 
\begin{eqnarray}
Q(x_0 =0,t) & = & 0 ,\\
\frac{\partial \ }{\partial x_0} Q(x_0 = L,t) & = & 0, 
\end{eqnarray}
for all $t$,  where $0\leq x_0 \leq L$.

To solve Eq.~(\ref{eq:backwardA1}), we 
apply the Laplace transform in the temporal variable, using $\mathcal{L}\{f(t)\}\equiv \tilde{f}(s)=\int_0^{\infty}f(t)e^{-t s}ds$, and additionally use the change of variables  
\begin{eqnarray}
 y(x)&=&\int_0^x D(x')^{-\frac{1}{2}}dx' ,
 \label{eq:maping}
\end{eqnarray}
which, to be well-defined, 
requires that the integrand not grow  faster than $x^2$ at the origin. In such case,  $y$ is a monotonically increasing function of $x$. 
The integral $y(x)$ will play a central role, as it contains the information about the profile $D(x)$, and,  the fact that its integrand depends only on $D(x)$  will print important properties to the heterogeneous search process.

After this change of variables, 
the solution in Laplace space is (for details, see   \ref{app:stratonovich}): 
\begin{eqnarray} \nonumber
\tilde{Q}(y_0,s)  &=& \frac{1}{s}\left( 1 - \frac{e^{2y_{L}\sqrt{s}-y_0\sqrt{s}}}{ e^{2y_{L}\sqrt{s}}+1}  - \frac{e^{y_0\sqrt{s}}}{e^{2y_{L\sqrt{s}}}+1}  \right)  \\
  &=& \frac{1}{s}\left(
1 -  \frac{\cosh( [y_L-y_0]\sqrt{s})}{\cosh(y_L\sqrt{s} )}
\right), 
\label{eq:QLaplaceStratA1}
\end{eqnarray}
where we defined $y_{L}\equiv y(L)$ and $y_0 \equiv y(x_0)$.
The Laplace transformed FPTD associated with the survival probability in Eq.~(\ref{eq:QLaplaceStratA1}) is given by

\begin{eqnarray}
\tilde{\wp}(y_0,s)  =  1-s\tilde{Q}(y_0,s)=
\frac{\cosh([y_L-y_0]\sqrt{s})}{\cosh(y_L\sqrt{s} )}.
 \label{eq:FIRSTpassage}
\end{eqnarray}
Using  Eqs. (\ref{eq:QLaplaceStratA1}) and (\ref{eq:FIRSTpassage}), we will be able to determine the measures $\langle 1/t \rangle$ and $1/\langle t \rangle$ for an arbitrary $D(x)$, implicitly embodied in  $y$.

\subsection{ Obtaining $1/\langle t \rangle$ 
} 
 
We first calculate the MFPT using  $\wp(t)=-\frac{d\ }{dt} Q(x_0,t)$, namely, 

\begin{eqnarray}
  \langle t \rangle &=&     \int_0^{\infty} t \,\frac{d\ }{dt} Q (x_0,t)  \, dt   
= \tilde{Q}(y_0,s\to 0). \label{eq:MFPT}
\end{eqnarray}
Thereby, using the form of $Q (x_0,t)$ in Eq. (\ref{eq:QLaplaceStratA1}), we have

\begin{eqnarray}
\langle t \rangle & = &   y_{L}y_0  - \frac{y_0 ^2}{2},
\end{eqnarray}
hence, 

\begin{eqnarray}
\langle t \rangle ^{-1}  =  
    \frac{2}{y_0( 2y_L- y_0)}, \label{eq:Ef2}
\end{eqnarray}
recalling that $y_L=y(L)$, where $y(x)$ was defined in Eq.~(\ref{eq:maping}).

On the other hand, if $D(x)$ does not grow faster than $x^2$ for large $x$, then, 
$L\to\infty$ implies  $ y_{L}\to \infty$. 
In such case we obtain that $\lim_{L\to \infty}\langle t \rangle^{-1} = 0$,
because when exploring the semi-infinite region,  $\langle t \rangle$ increases indefinitely~\cite{redner2001guide}.  

\subsection{ Obtaining $\langle 1/t \rangle $ }

The average $\langle 1/t \rangle$ defined in Eq.~(\ref{eq:inverset}) can be rewritten in terms of the FPTD  (\ref{eq:FIRSTpassage}) in Laplace space, as follows

\begin{eqnarray}
 \langle t^{-1} \rangle & = & 
 \int_0^\infty  t^{-1}\, \wp(y_0,t)\, dt  
 = \int_0^\infty \tilde{\wp}(y_0,s)\, ds 
 \nonumber \\
& = &  \int_0^\infty 
\frac{\cosh([y_L-y_0]\sqrt{s})}{\cosh(y_L\sqrt{s})}ds. \label{eq:mesure1Strat}
\end{eqnarray}

Defining the new variable $z= e^{2y_{L}\sqrt{s}}$, Eq.~(\ref{eq:mesure1Strat}) becomes

\begin{eqnarray} \nonumber
  \langle t^{-1}\rangle &=&  \int_1^{\infty} \frac{ z^{-\frac{1}{2}\frac{y_0}{y_{L}}} + z^{\frac{1}{2}\frac{y_0}{y_{L}}-1} }{2 y_{L}^2 }  \frac{\log z}{z+1}   dz  \\ \nonumber
    &= & \frac{1}{8 y_{L}^2 } \left[\psi ^{(1)}\left(\frac{y_0/y_{L}}{4}\right)-\psi ^{(1)}\left(\frac{y_0/y_{L}+2}{4}\right)  \right.  
    \\ & & - \left. \psi ^{(1)}\left(1-\frac{y_0/y_{L}}{4}\right)+\psi ^{(1)}\left(\frac{1}{2}-\frac{y_0/y_{L}}{4} \right) \right],   
    \;\;\; \label{eq:Ef1}
\end{eqnarray}
with $y_0/y_L \in [0,1]$, and where  $\psi^{(1)}(z)$ is the polygamma function~\cite{abramowitz1965handbook}.

\subsection{Properties for $x_0=L$}
\label{sec:x0=L}

 Note that both $1/\langle t \rangle$ and $\langle 1/t \rangle$ 
depend on the diffusivity profile through $y_0$ and $y_L$ only, which integrate a function of the diffusivity from the target to $x_0$ and up to $L$ respectively. This implies that, curiously,  shuffling the values of the profiles within each one of the intervals $(0,x_0)$  and $(x_0,L)$  will not alter the results.

\begin{figure}[h!]  
\centering
\includegraphics[width=0.6\textwidth]{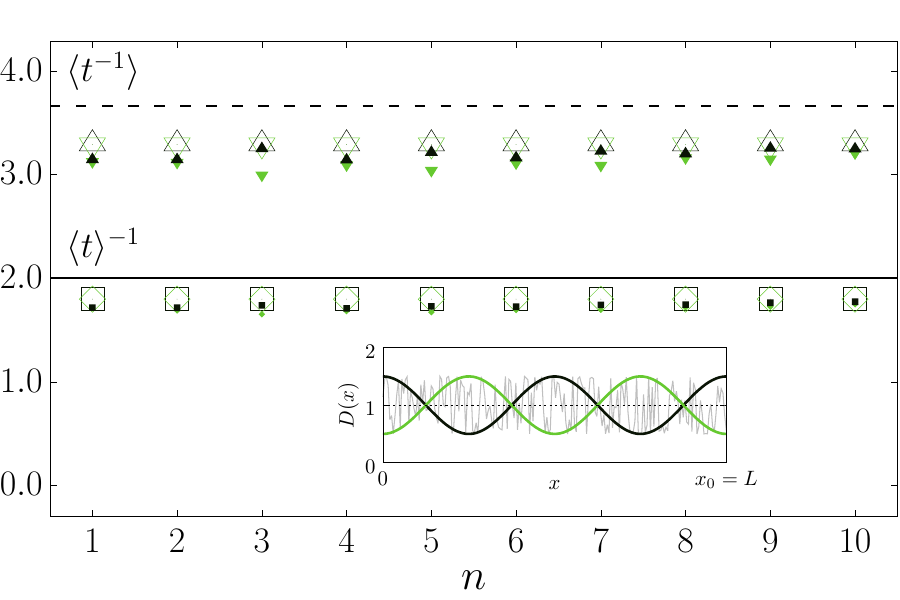} 
\caption{Within Stratonovich framework, we plot $1/\langle t\rangle$ (large triangles)  and  $\langle1/t\rangle$ (large quadrangles), using Eqs. (\ref{eq:Ef2}) and   (\ref{eq:Ef1}), respectively,  
for the diffusivity profiles of the form 
$D(x)=1 + d\cos(n\pi x/L)$, with $d=\pm 0.5$ (illustrated in the inset for $n=4$) as a function of integer values of $n$. 
The small symbols represent  the corresponding averages over $10^6$ realizations obtained from the integration of the stochastic simulation of Eq.~(\ref{eq:langevin2}), 
using the Euler method with a time step $10^{-5}\le \Delta t \le 10^{-3}$.
In the inset, we included a shuffled profile (after fragmentation in 200 segments), for which the same invariant levels are obtained. 
The horizontal lines (solid for $1/\langle t\rangle$ and  dashed for $\langle1/t\rangle$)
represent the values for the homogeneous case $d=0$, which performs better than the heterogeneous case, according to both measures.   
 }
\label{fig:sinoidal}
\end{figure}

The particular case $x_0 = L$  means that the searcher is initially positioned  on the reflecting wall,  which implies $y_{L}=y_0$. In such case, the results are not altered by shuffling the values of the diffusivity in the whole accessible region. Similar property was observed for the unbounded case ($L\to \infty$)~\cite{Santos2022}. 
In our case, the effect is illustrated in Fig.~\ref{fig:sinoidal}, where we consider the family of profiles 
$D(x)=1+d\cos(n\pi x/L)$, with $n=1,2,\ldots,10$, and $d=\pm0.5$, which all contain the same values with equal probability (as can be seen by stretching and unfolding). In fact, for any integer $n$, the theoretical result, for each efficiency (large symbols), remains invariant, in good agreement with the simulational data (small symbols). 
Furthermore, we verified that for the shuffled profile shown  in the inset (after decomposition in 200 fragments), Langevin simulations also yield the expected same level. 

Let us remark that, changing the average level $D_0$ of the diffusivity proportionally changes the efficiency measures, then, for fair  comparisons in all numerical examples, we set unitary average  level.

Moreover, for $x_0=L$, hence $y_0=y_L$, Eq.~(\ref{eq:Ef2}) becomes
\begin{eqnarray}
\langle t \rangle ^{-1}\Big|_{y_0= y_{L}}  =  
    \frac{2}{y_L^2}. 
    \label{eq:Ef2=} 
\end{eqnarray}
While, using Eq.~(\ref{eq:mesure1Strat}) and the 
the change of variables $z=e^{y_{L}\sqrt{s}}$, we obtain   
\begin{eqnarray}
\langle  t^{-1} \rangle\Big|_{y_0= y_{L}} & = &  \int_0^\infty \frac{ 1}{\cosh(y_L\sqrt{s})}ds = \frac{4}{y_L^2} \int_1^{\infty}  \frac{ \log z}{ z^2+1}  dz =  \frac{4G}{y_L^2}  \nonumber \\
&\simeq & 1.83 \times \frac{2}{y_L^2},
\label{eq:Ef1=}
\end{eqnarray}
where $G\simeq 0.915$ is the Catalan's constant.
Therefore, a general proportionality exists between both measures independently of $D(x)$, when the searcher is initially at the wall ($x_0=L$). 

The limit $L\to \infty$, implying  $y_L  \to \infty$, when applied in Eq.~(\ref{eq:FIRSTpassage}), leads to  $\tilde{\wp}(y_0,s)  =  e^{-y_0\sqrt{s}}$, which substituted in the first line of  Eq.~(\ref{eq:mesure1Strat}) immediately yields
\begin{eqnarray}
\langle  t^{-1} \rangle \Big|_{ y_{L}\to \infty }  & = &  \frac{2}{y_0^2}.  \label{eq:Ef1infinity}
\end{eqnarray}
This result was already found in Ref.~\cite{Santos2022},  where  semi-infinite environment was investigated. 
For such unbounded domain,  
particles far from the target will likely reach it in so long times that will have a negligible contribution to $\langle t^{-1}\rangle$.
Here we remark the coincidence of Eq.~(\ref{eq:Ef1infinity}) with 
Eq.~(\ref{eq:Ef2=}) when $x_0=L$. 
Results are summarized in Table~\ref{tab:table}. 
\begin{table}[ht]
\centering
\begin{tabular}{c|c|c|}
\cline{2-3}
 & $x_0=L$  & $L \to \infty$ \\ \hline
\multicolumn{1}{|l|}{$ 1/\langle t \rangle$} &  $2/ y_L^{2}$   &  0  \\ \hline
\multicolumn{1}{|l|}{ $ \langle 1/t \rangle$  }  & $\sim 1.83 \times 2/ y_L^{2}$     &  $2/y_0^{2}$  \\ \hline 
\end{tabular}
\caption{Comparison of particular cases within Stratonovich framework. Recalling that $(0,L]$ is the  accessible region, $x_0$ the initial position of the searcher for a target at $x=0$, and $y(x)$ is given by Eq.~(\ref{eq:maping}).}
\label{tab:table}
\end{table}

Given the inverse scaling of the efficiency with $y_L$, when $x_0=L$, 
another important property emerges within the Stratonovich framework. 
 On the one hand,  it can be shown~\cite{dos2020critical} that 
 \begin{equation} \label{eq:YLma}
 y_L=\int_0^L D(x)^{-\frac{1}{2}}dx \ge LD_0^{-1/2},
 \end{equation}
 where $D_0$ is the average level  of the diffusivity in $[0,L]$.
Since the efficiencies scale inversely with $y_L$, 
then the inequality (\ref{eq:YLma}) means that in an   
heterogeneous (nonH) profile the searcher performs less efficiently than in an homogeneous (H) environment with average level. That is, we find as a general result, valid for the Stratonovich framework when $x_0=L$, that
the efficiency $\varepsilon$,  measured either by $1/\langle t \rangle$ or $1/\langle t \rangle$, verifies
 \begin{equation} \label{eq:epsilon}
 \varepsilon_{nonH} \le \varepsilon_{H}. 
 \end{equation}
This effect is   illustrated  in  Fig.~\ref{fig:sinoidal}, where the corresponding homogeneous cases are represented by horizontal lines, above the heterogeneous values.

\subsection{Properties for arbitray initial position $x_0 \in (0,L]$}
\label{sec:x0<L}

%%When $x_0<L$, the shuffling property applies separately to the segments $[0,x_0]$ and $[x_0,L]$. 

\begin{figure}[h!]  
\centering
 \includegraphics[width=0.49\textwidth]{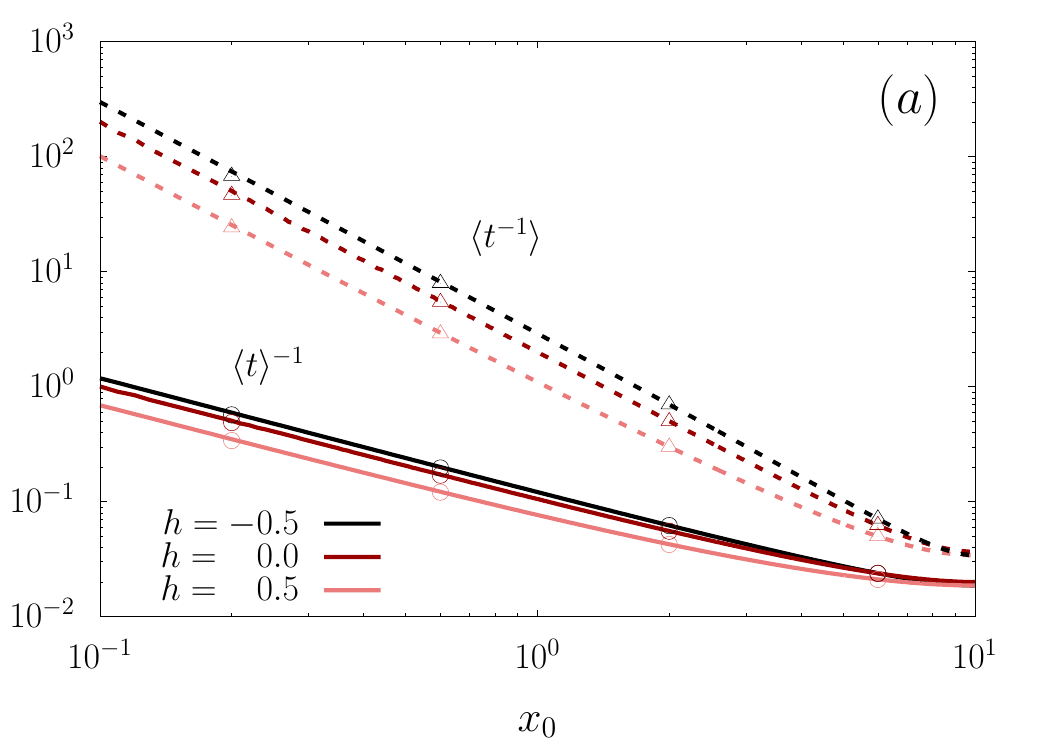} 
\includegraphics[width=0.49\textwidth]{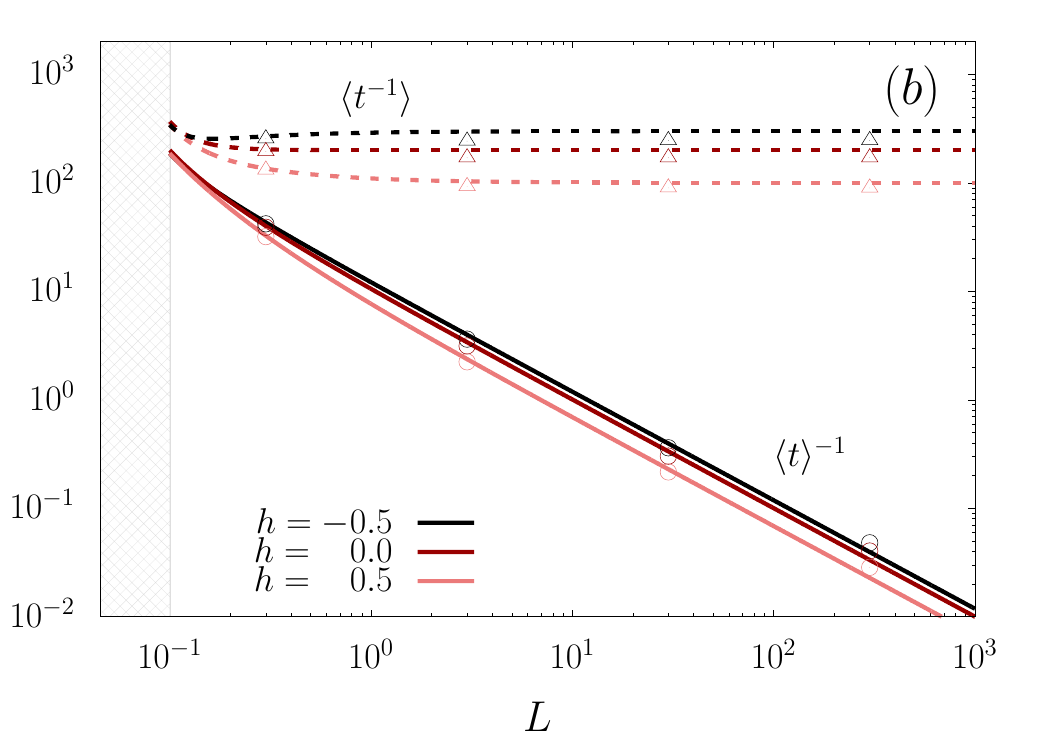}  
\caption{ $1/\langle t\rangle$ (solid lines)  and  $\langle1/t\rangle$ (dashed lines), using Eqs. (\ref{eq:Ef2}) and   (\ref{eq:Ef1}), 
for the linear diffusivity profile 
%$D(x)= x^\alpha$, for different values of  $\alpha$. 
%
$  D(x)= 1 + h(2x/L-1)$,  for different values of $h$,   
as a function of  the initial position $x_0$, fixing $L=10$ (a) and as a function of the boundary position $L$, fixing $x_0=0.1$ (b). 
The symbols represent  the stochastic simulation of Eq.~(\ref{eq:langevin2}), averaged over $10^6$ realizations. 
}
\label{fig:vsx0}
\end{figure}

 As paradigmatic example we first consider the linear profile  
  $D(x)= 1 + h(2x/L-1)$, which 
encompasses diffusivity profiles that increase ($0<h<1$) or decrease ($-1<h<0$) from the position of the target, as well as the homogeneous case ($h=0$). Note that, if $h>0$ (resp. $<0$), the diffusivity increases (resp. decreases) when moving away from the target). 
 Moreover this family has fixed (unitary) average level, which turns comparisons  fair. 
For these linear profiles, 
we plot, in Fig.~\ref{fig:vsx0}, the theoretical results for $\langle1/t\rangle$   and   $1/\langle t\rangle$ vs. $x_0$ (a) and vs. $L$ (b),  in good agreement with the simulations.  
In panel (a), we can observe that, besides decaying with the distance $x_0$, as expected, both measures approach each other when  $x_0\to L$, reaching the ratio $2G$ when $x_0=L$, as predicted by Eqs.~(\ref{eq:Ef2=}) and (\ref{eq:Ef1=}).

These effects can be also observed (not shown) for the power-law family 
 $D(x)=x^\alpha$, where $\alpha >(<)0$ also represents profiles that 
 increase (decrease) from the target, although in this case the 
 average level varies with $\alpha$.

With regard to the dependence on $L$, 
$1/\langle t\rangle$   decays with $L$, since  
 the MFPT grows with $L$, indicating that individual arrival times also grow, however, these times do not have a significant contribution to $\langle1/t\rangle$, which for large $L$ tends to a constant. 
 This dichotomy  reflects the fact that for large $L$ both efficiencies measure different properties, to the extent that long search times become more likely.  
For both measures, the case $h<0$ (implying larger diffusivity near the target) allows a better performance

\section{Search efficiency under arbitrary interpretation of the HDP}
\label{sec:all}

The backward FP equation for all interpretations of the HDP  is given by
 
\begin{eqnarray} 
\frac{\partial \ }{\partial t} Q(x_0,t) & = &    D(x_0)^{\frac{A}{2}}\frac{\partial \ }{\partial x_0}\left\{D(x_0)^{1-\frac{A}{2}} \frac{\partial \ }{\partial x_0}  Q(x_0,t) \right\}\,,\label{eq:backwardA}
\end{eqnarray} 
recalling that $Q(x_0,t)$ is the survival probability and $x_0$ the initial position. 

\subsection{MFPT for arbitrary $D(x)$}
\label{sec:mfpt}

According to Eq.~(\ref{eq:backwardA}) the mean first-passage time, i.e., $\langle t \rangle$,  is given by
\begin{eqnarray} 
\langle t \rangle  & = & \int_0^{L} D(x'')^{-\frac{A}{2}}dx''  \int_0^{x_0} D(x')^{-1 + \frac{A}{2}}dx'  \nonumber \\  & - &  \int_0^{x_0} D(x'')^{-1 + \frac{A}{2}}  \int_0^{x''} D(x')^{-\frac{A}{2}}dx' dx'' ,
\label{eq:GeneralT}
 \end{eqnarray}
 whose derivation can be found  in  \ref{app:mfpt-anyA}. 
 From Eq.~(\ref{eq:GeneralT}), it  is possible to calculate $1/ \langle t \rangle$ for all the HDP interpretations. 
 In the case $A=1$ (Stratonovich), for $\langle t \rangle$, we recover  Eq.~(\ref{eq:Ef2}).

With this general expression for the  MFPT,  we can study the implications of It\^o 
and anti-It\^o interpretations. 

Considering $x_0=L$ in Eq.~(\ref{eq:GeneralT}), for $A=0$ and $A=2$, we have
\begin{eqnarray}\label{eq:A0}  
  \langle t\rangle_{A=0}   & = &  \int_0^{L} 
  \frac{L-x}{D(x)}   dx  
%%%& = &  -\int_{x_0}^{0}  \frac{x'}{D_1(x_0-x')} dx'  \nonumber \\
= \int^{L}_{0} \frac{x'}{D(L-x')} dx' 
\end{eqnarray}
and 
\begin{eqnarray}\label{eq:A2}
 \langle t\rangle_{A=2}   & = & \int_0^{L} \frac{x}{D(x)}dx.
\end{eqnarray}

First we remark that, in contrast to Stratonovich framework ($A=1$), the integrands in Eqs.~(\ref{eq:A0}) and (\ref{eq:A2}) 
do not depend only on $D$.   
This means that the insensitivity to the ordering of the heterogeneity, observed for the MFPT when  $A = 1$, and $x_0=L$, is broken for other interpretations of the HDP, implying that  the shape of the profile is relevant and not only the distribution of values within the accessible domain. 

\begin{figure}[ht]  
\centering
\includegraphics[width=0.49\textwidth]{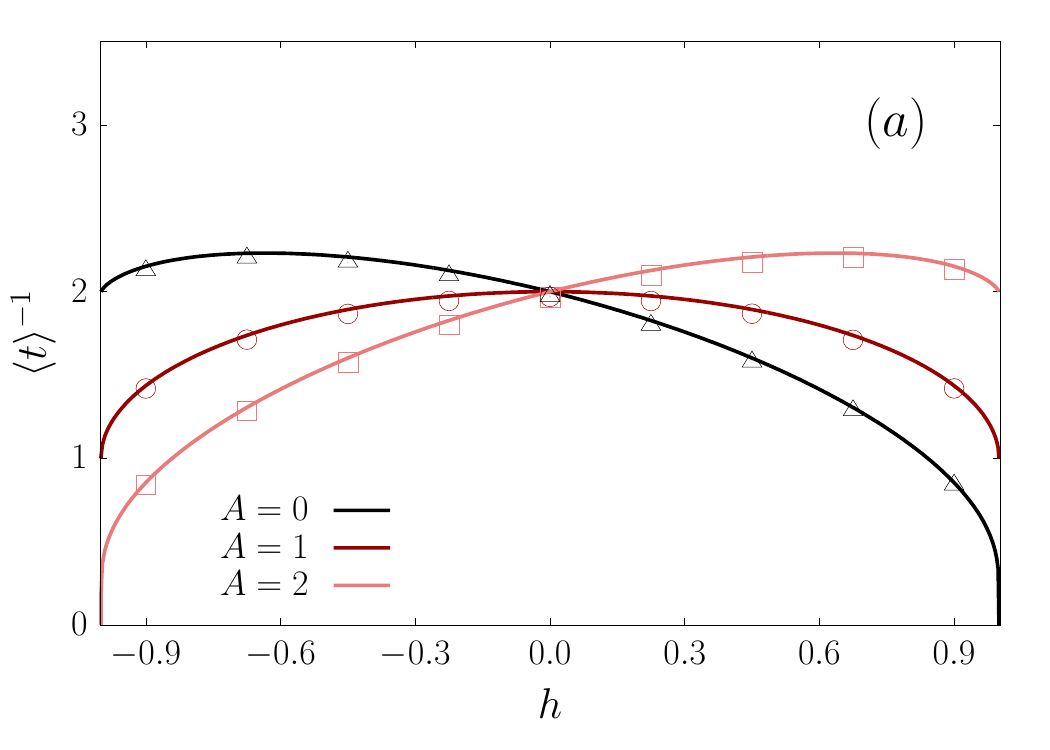} 
\includegraphics[width=0.49\textwidth]{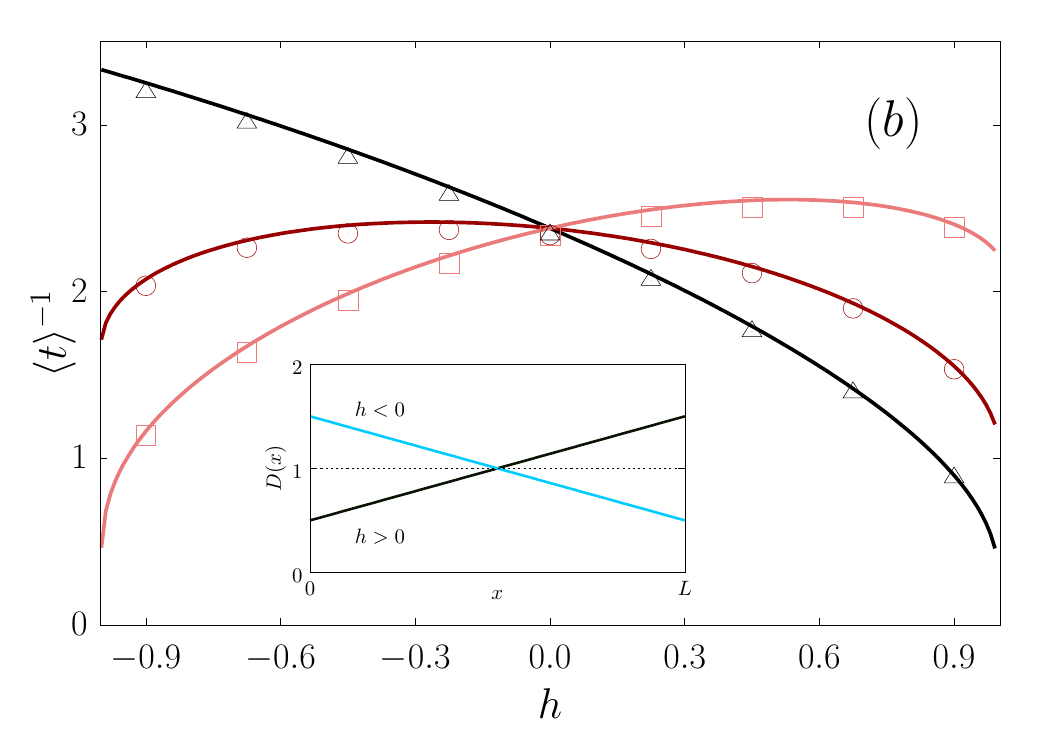} 
\caption{ Efficiency $\langle t \rangle ^{-1}$,  obtained from Eq.~(\ref{eq:GeneralT}) (solid lines) and from numerical simulations (symbols), for the diffusivity profile %$D(x)=1+d\cos(\pi x/ L)$, 
$D(x)=1+h(2x/L-1)$ (depicted in the inset of panel (b)),  for 
 different values of $A$, setting $L=1$,  and 
(a) $x_0=1$ and (b) $x_0=0.6$. 
Notice in panel (a) that since $x_0=L$, exchanging  $x \leftrightarrow L-x$ is equivalent to the change of interpretation parameter $A$: $0 \leftrightarrow 2$, while this symmetry is broken if $x_0<L$, as illustrated in panel (b).  
}  
\label{fig:Ef1/<t>}
\end{figure}

 Second, note that  It\^o and anti-It\^o interpretations produce the same 
outcomes for profiles that are symmetric around $L/2$, when $x_0=L$. 
That is, the reflection of the heterogeneity profile  $D(x) \leftrightarrow  D(L-x)$ emulates the anticipating It\^o ($A=0$) and  non-anticipating anti-It\^o ($A=2$) character of the stochastic integration. 
This symmetry effect is illustrated in Fig.~\ref{fig:Ef1/<t>}(a), using the linear profile $D(x)= 1 + h(2x/L-1)$, 
 which is monotonic in the accessible region $[0,L]$, and setting $x_0=L$. 
Notice, however that this symmetry does not  hold if $x_0<L$, 
as can be seen in Fig.~\ref{fig:Ef1/<t>}(b).
Qualitatively similar results are obtained for  nonlinear profiles, e.g., for 
$D(x)=  1+d\cos( \pi x/L) $ (not shown).

\begin{figure}[ht]  
\centering
\includegraphics[width=0.49\textwidth]{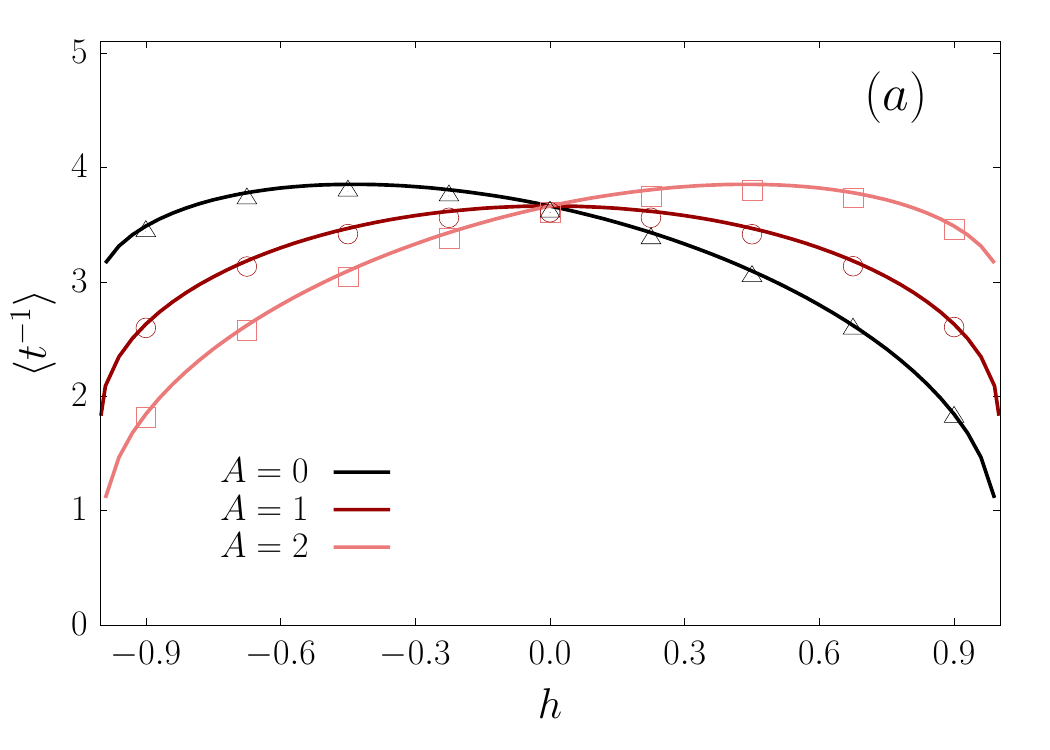} 
\includegraphics[width=0.49\textwidth]{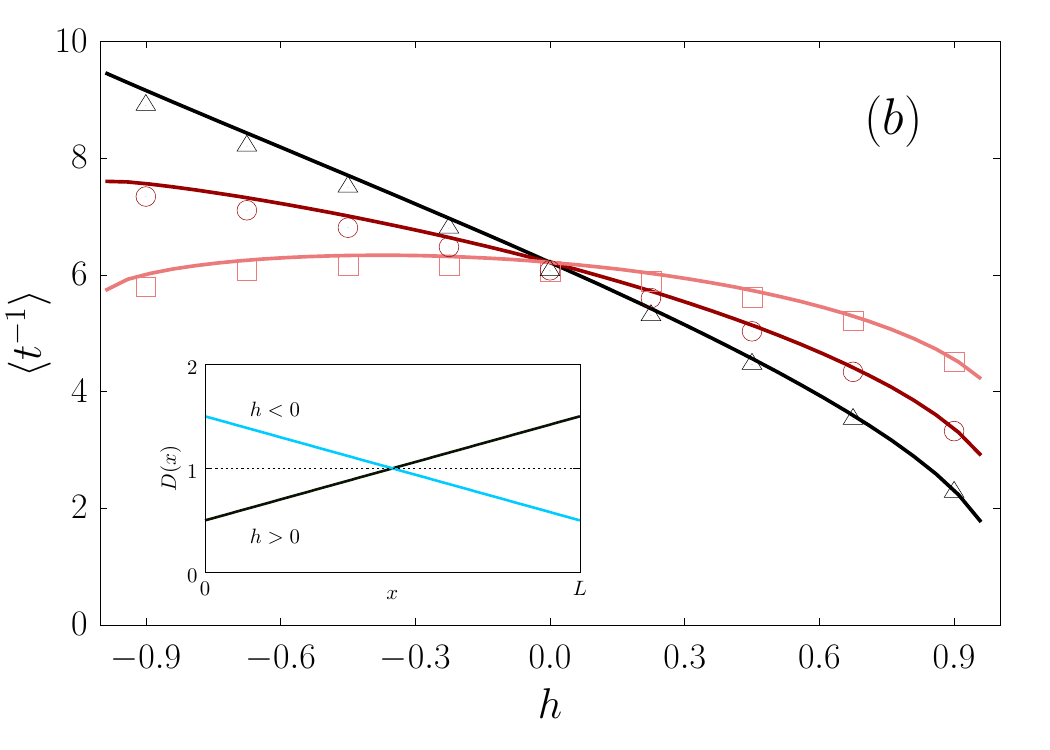} 
\caption{ Efficiency $\langle t^{-1} \rangle$ obtained from Eq.~(\ref{eq:ps}) (solid lines) and from numerical simulations (symbols), 
with the same diffusivity and parameters used in Fig.~\ref{fig:Ef1/<t>}.
}  
\label{fig:Ef<1/t>}
\end{figure}

It is important to note, that this exchange symmetry is already present in  Langevin Eq.~(\ref{eq:langevin2}). 
Indeed, if we use It\^o's Lemma for the linear change of variables $x=L-\bar{x}$, together with the exchange $A=2-\bar{A}$, the Langevin equation remains the same. But the observed exchange symmetry emerges only when the initial position is at the reflecting boundary $x_0=L$. 
Then, similar behaviors are expected for  $\langle 1/t \rangle$, 
as seen for the linear profile in 
Fig.~\ref{fig:Ef<1/t>}, which will be developed in the next section.

Another important implication of the results for $x_0=L$ is that while for the Stratonovich interpretation the searcher performs better in homogeneous media,  for $A\neq 1$  there is a heterogeneous profile that optimizes the search, although it is not the same for both efficiencies. 

Finally, we observe that, when the profile increases with the
distance from the target $(h > 0)$, the efficiency increases with larger $A$, while the contrary occurs for a decreasing profile $(h< 0)$. This effect has been previously observed and explained for the unbounded domain in Ref.~\cite{Santos2022}. Differently, when $A=1$, the shape of the profile is not relevant, but only its distribution of values, hence the results for given $h$ coincide with those for $-h$.

\subsection{ Search efficiency $\langle 1/t \rangle$ for a special family of profiles $D(x)$}
\label{sec:power}

Although the next calculations can be straightforwardly extended to the family of profiles of the power-law form 
 $D(x) =  (a x + b)^\alpha $, such that $ax+b>0$ for  $x\in [0,L]$ and $\alpha<2$, we will develop explicitly the linear case, setting $\alpha=1$, which provide illustrative examples with less complicated expressions. 
 
Starting from Eq.~(\ref{eq:backwardA}),  
in \ref{app:ax+b}),  we obtain the FPTD in Laplace space

\begin{eqnarray} \nonumber
\tilde{\wp}(s) &= &   
\frac{  (a x_0+b)^\frac{\gamma}{2} }{ b^\frac{\gamma}{2}} \; 
\frac{
I_{\gamma-1} \left[ \xi(s,L) \right] \, K_{\gamma} \left[ \xi(s,x_0) \right]  + 
   K_{\gamma  -1 } \left[ \xi(s,L) \right]\, I_{\gamma } \left[ \xi(s,x_0) \right] 
}{   I_{\gamma -1 }\left[\xi(s,L) \right]  K_{\gamma} \left[ \xi(s,0)\right]+   K_{\gamma -1 }\left[ \xi(s,L)\right]  I_{\gamma } \left[ \xi(s,0) \right] }, \\ \label{eq:ps}
\end{eqnarray}
where $\gamma =  A/2$ and $\xi(s,x_0) = 2\sqrt{s(a x_0+b)} /a$.  

Then, we can calculate $\langle 1/t\rangle$ by numerical integration of equation 
$\left\langle 1/t \right\rangle =  \int_0^{\infty} \tilde{\wp}(s) ds$.
Results, compared to Langevin simulations, are displayed in Fig.~\ref{fig:Ef<1/t>} for the linear profile with unitary average, 
 $D(x)=1+h(2x/L-1)$. 
 Note that the symmetries and main features discussed   for $1/\langle t \rangle$, when $x_0=L$,  also hold for $1/\langle t \rangle$.

\section{Final remarks}
\label{sec:final}

We have calculated and compared  two standard measures of efficiency for the random search problem in a one-dimensional bounded interval where the diffusion coefficient is space-dependent. 
 
The first step was to consider a general  profile $D(x)$ under the Stratonovich interpretation ($A=1$), for which we discussed  the similarities and differences between both efficiencies. 
A relevant common feature is that they do not depend on the sequence of values of $D(x)$, since the dependence on the diffusivity profile occurs only through the integral $y$ of a function of $D(x)$. 
Particular features emerge when $x_0=L$ (search initialized on the reflecting wall), namely, we have shown that
(i) heterogeneous profiles are less efficient that the homogeneous one with same average diffusivity
and (ii) there is a proportionality    
between both measures, which is valid regardless of the profile $D(x)$.

For general interpretation of stochastic integration, characterized by parameter $A$,  
we obtained  results for $1/\langle t \rangle$,  valid for any $D(x)$. From which, an important consequence is a symmetry property that emerges when $x_0=L$, namely, the change  $D(x) \leftrightarrow D(x-L)$ yields the same results when  changing the interpretation parameter $A$: $0 \leftrightarrow 2$. 
This symmetry also emerges for the measure of efficiency  $ \langle 1/t \rangle$, for which we managed to obtain an analytical expression for particular choices of $D(x)$, as 
 the  power-law diffusivity, among which we developed the linear case, which embraces environments with increasing and decreasing mobility versus the distance from  the target, as well as the homogeneous case. 
 Furthermore,  in contrast to the case   $A=1$, for which the homogeneous profile with equal average diffusivity allows a more efficient search than heterogeneous ones, when $A\neq 1$,   heterogeneous environments can enhance the efficiency to reach the target. 
Another general feature is that, increasing $A$ favors  the search when the   diffusivity increases with the distance from the target and hinders the search otherwise.
Moreover, this is not unique to a particular shape of the diffusivity profile, or initial position, but is determined by its monotonic character.

 As  perspectives of continuation, it would be interesting to extend the present study to higher dimensions, consider colored instead of white noise, introduce stochastic resetting, among other variants.

{\bf Acknowledgments:} 
We all acknowledge partial financial support by the 
Coordena\c c\~ao de Aperfei\c coamento de Pessoal de N\'{\i}vel Superior
 - Brazil (CAPES) - Finance Code 001. C.A. also acknowledges partial support by 
 Conselho Nacional de Desenvolvimento Cient\'{\i}fico e Tecnol\'ogico (CNPq), 
and Funda\c c\~ao de Amparo \`a Pesquisa do Estado do Rio de Janeiro (FAPERJ).

\bibliographystyle{iopart-num}
\section*{References}
%\bibliography{ref}
 \providecommand{\newblock}{}

\newpage

\appendix

\section{FPTD in Laplace space for Stratonovich case}
\label{app:stratonovich}

Using the change of variables defined in Eq.~(\ref{eq:maping}),  
 for the new variable $y$,  
 Eq.~(\ref{eq:backwardA1}) becomes
\begin{equation}
\frac{\partial}{\partial t} Q(y_0,t) = \frac{\partial^2}{\partial y_0^2} Q(y_0,t).
\end{equation}
Upon performing a Laplace transform with respect to time, the equation becomes
\begin{equation} \label{eq:FP-A=1}
s \tilde{Q}(y_0,s) - 1 = \frac{\partial^2}{\partial y_0^2} \tilde{Q}(y_0,s).
\end{equation}
This leads to
\begin{equation}
\tilde{Q}(y_0,s) = \frac{1}{s}\left( 1 + B_1(s) e^{-y_0\sqrt{s}} + B_2(s) e^{y_0\sqrt{s}} \right).
\label{eq:suvivorAB}
\end{equation}
Applying the boundary conditions $Q(0,s)=0$ and $\left.\partial_{y}Q\right|{y_L}=0$, we obtain
\begin{eqnarray} \nonumber
B_1(s) & = &- B_2(s) - 1 = e^{2y{L}\sqrt{s}} B_2(s) \,.
\end{eqnarray}
Solving for $B_1(s)$ and $B_2(s)$, gives
\begin{eqnarray} \nonumber
B_1(s) & =& \frac{-e^{2y_{L}\sqrt{s}}}{ e^{2y_{L}\sqrt{s}}+1}, \\  \nonumber
B_2(s) & =& \frac{-1}{e^{2y_{L}\sqrt{s}}+1}. 
\end{eqnarray}
Substituting these results back into Eq.~(\ref{eq:suvivorAB}), we get Eq.~(\ref{eq:QLaplaceStratA1}).

\section{MFPT  for arbitrary $A$ } 
\label{app:mfpt-anyA}

The Laplace transform of Eq. (\ref{eq:backwardA}) is  
\begin{eqnarray} \label{eq:D1}
s \tilde{Q}_{s}(x_0) - 1 & = &   D(x_0)^{\frac{A}{2}} \frac{\partial \ }{\partial x_0} \left\{ D(x_0)^{1-\frac{A}{2}} \frac{\partial \ }{\partial x_0}  \tilde{Q}_{s}(x_0) \right\}. 
\end{eqnarray}
By defining the MFPT  $T(x_0) = \langle t \rangle = \lim_{s\to 0} \tilde{Q}_{s}(x_0)  $,   Eq.~(\ref{eq:D1}) implies
\begin{eqnarray} \nonumber     -  D(x_0)^{-\frac{A}{2}}   & = &
\frac{\partial \ }{\partial x_0} \left\{ D(x_0)^{1-\frac{A}{2}} \frac{\partial \ }{\partial x_0}  T(x_0) \right\} .
\end{eqnarray}
After the first integration in the variable $x_0$, we obtain
\begin{eqnarray} \nonumber
\frac{\partial \ }{\partial x_0}  T(x_0)  & = & B  D(x_0)^{-1 + \frac{A}{2}}  -D(x_0)^{-1 + \frac{A}{2}}  \int_0^{x_0} D(x')^{-\frac{A}{2}}dx'\,,
\end{eqnarray}
where $B$ is a  constant. After the second integration in $x_0$, we get
\begin{eqnarray} 
 T(x_0)  & = & B  \int_0^{x_0} D(x')^{-1 + \frac{A}{2}}dx'   \nonumber \\ &- & 
 \int_0^{x_0} D(x'')^{-1 + \frac{A}{2}}  \int_0^{x''} D(x')^{-\frac{A}{2}}dx' dx''. 
 \label{eq:Tx0} 
\end{eqnarray}
To find out the constant $B$, we consider  the reflection condition 
 (that is, $\partial_{x_0} \tilde{Q}|_{x_0=L}=\partial_{x_0} T|_{x_0=L} =0$, for $s\to 0$), then   
\begin{eqnarray} \nonumber
 B  D(L)^{-1 + \frac{A}{2}} -  D(L)^{-1 + \frac{A}{2}}  \int_0^{L} D(x')^{-\frac{A}{2}}dx' =0,
\end{eqnarray}
that leads to
\begin{eqnarray}
 B & = & \int_0^{L} D(x')^{-\frac{A}{2}}dx'. 
\end{eqnarray}
Substituting $B$ into Eq.~(\ref{eq:Tx0}), 
we arrive to Eq.~(\ref{eq:GeneralT}).

 \section{Survival 
probability for $D(x) = a x + b$}
\label{app:ax+b}

Applying  the transformation defined in Eq. (\ref{eq:maping}) into Eq. (\ref{eq:backwardA}),  we obtain 
\begin{eqnarray} 
\frac{\partial \ }{\partial t} Q(y_0,t) & = & \frac{1-A}{2\sqrt{D(x)}} \left.\frac{d D(x)}{dx}\right|_{y_0} \frac{\partial \ }{\partial y_0} Q(y_0,t) + \frac{\partial^2 \ }{\partial y_0^2} Q(y_0,t)
\,, \label{eq:diff1}
\end{eqnarray}
which for $A=1$ recovers Eq.~(\ref{eq:FP-A=1}).

For a linear profile $D(x)= a x+b$, 
 Eq.~(\ref{eq:maping}) becomes $y(x)=2(\sqrt{D(x)}-\sqrt{b})/a$. 
 Then Eq.~(\ref{eq:diff1}) can be rewritten as 
\begin{eqnarray} 
\frac{\partial \ }{\partial t} Q(y_0,t) & = & \frac{1-A}{\mu +y_0}\frac{\partial \ }{\partial y_0} Q(y_0,t) + \frac{\partial^2 \ }{\partial y_0^2} Q(y_0,t)
\,,\label{eq:diff2}
\end{eqnarray}
where $\mu=2\sqrt{b}/a$. 

The  Laplace transform of Eq.~(\ref{eq:diff2}) is given by 
\begin{eqnarray} 
\frac{\partial^2 \ }{\partial y_0^2} \tilde{Q}(y_0,s) +   \frac{1-A}{\mu +y_0}\frac{\partial \ }{\partial y_0} \tilde{Q}(y_0,s) -s \tilde{Q}(y_0,s) + 1 = 0
\,,\label{eq:backwarddif2}
\end{eqnarray}
and using the  ansatz
\begin{equation} \label{eq:Q}
    \tilde{Q}(y_0,s)= (y_0 + \mu)^{\frac{A}{2}}R(y_0,s) + 1/s \,,
\end{equation}
we rewrite Eq. (\ref{eq:backwarddif2}) as
\begin{equation} \nonumber
    \frac{\partial^2}{\partial y_0^2}R(y_0,s) +\frac{1}{(y_0 +\mu)}\frac{\partial}{\partial y_0}R(y_0,s) - \left[ \left(\frac{A}{2(y_0 +\mu)}\right) + s \right]R(y_0,s) =0, 
\end{equation}
which can be identified as a generalization of the Bessel equation 
of order $A/2$~\cite{bowman2012introduction}, whose solution is the linear combination
\begin{equation}
    R(y_0,s)= c_1 K_{\frac{A}{2}}\left[\sqrt{s}(y_0+\mu)\right] +  c_2 I_{\frac{A}{2}}\left[\sqrt{s}(y_0+\mu)\right].
\end{equation}
From Eq.~(\ref{eq:Q}), we have
    \begin{equation} \label{eq:Qy}
    \tilde{Q}(y_0,s) = (y_0+\mu)^{\frac{A}{2}}\left(c_1 K_{\frac{A}{2}}\left[\sqrt{s}(y_0+\mu)\right] +  c_2 I_{\frac{A}{2}}\left[\sqrt{s}(y_0+\mu)\right] \right) +\frac{1}{s} \,. 
\end{equation}
The boundary condition $\tilde{Q}(0,s)=0$   gives 
\begin{equation}
    c_1 = - \frac{\mu^{-\frac{A}{2}}}{s K_{\frac{A}{2}}(\mu\sqrt{s})} -  c_2\,\frac{ I_{\frac{A}{2}}(\mu\sqrt{s})}{ K_{\frac{A}{2}}(\mu\sqrt{s})}  ,
    \label{eq:c1}
\end{equation}
and  the reflection boundary condition, $\left.\frac{\partial Q(y_0,s)}{\partial y_0}\right|_{y_0=y_L}=0$, gives
\begin{eqnarray} \nonumber
    c_2 &=& c_1 \frac{K_{\frac{A}{2}-1}[(y_L +\mu)\sqrt{s}]}{I_{\frac{A}{2}-1}[ (y_L +\mu)\sqrt{s}]}.
    \label{eq:c2}
\end{eqnarray}
Combining  Eqs.~(\ref{eq:c1}) and (\ref{eq:c2}), we obtain 
\begin{eqnarray} \nonumber
    c_1 &=&-\frac{1}{\mu^{A/2} s }\;\frac  { I_{\frac{A}{2}-1}\left[(y_L+\mu) \sqrt{s} \right] 
    }{ 
  I_{\frac{A}{2}-1}\left[ (y_L+\mu)\sqrt{s}\right] 
   K_{\frac{A}{2}}\left[ \mu \sqrt{s}\right]
   +
  K_{\frac{A}{2}-1}\left[  (y_L+\mu)\sqrt{s}\right] 
  I_{\frac{A}{2}}\left[  \mu \sqrt{s}\right]}.
\end{eqnarray}

Finally, inserting Eq.~(\ref{eq:Qy}) into  the relation $\tilde{\wp}(s) = 1 - s \tilde{Q}(s)$, we obtain the FPTD in Laplace space
\begin{eqnarray} \label{eq:Ps}
    \wp(s) &=& 
    \frac{ (y_0+\mu)^{\frac{A}{2}}}{\mu ^{A/2}}
    \frac{ 
    I_{\frac{A}{2}-1}\left[ y_L^\prime \sqrt{s} \right]
    K_{\frac{A}{2}}\left[ y_0^\prime\sqrt{s}\right]  
    +  K_{\frac{A}{2}-1}\left[y_L^\prime\sqrt{s} \right] 
    I_{\frac{A}{2}}\left[ y_0^\prime\sqrt{s}\right]}
    { I_{\frac{A}{2}-1}\left[  y_L^\prime\sqrt{s}\right] 
   K_{\frac{A}{2}}\left[ \mu \sqrt{s}\right]
   +
  K_{\frac{A}{2}-1}\left[ y_L^\prime\sqrt{s}\right] 
  I_{\frac{A}{2}}\left[  \mu \sqrt{s}\right]},
\end{eqnarray} 
where we defined $y_0^\prime \equiv y_0+\mu$ and 
$y_L^\prime \equiv y_L+\mu$. 
Recalling  the change of variables used in Eq.~(\ref{eq:diff2}), $y(x)=2\sqrt{ax+b}/a-\mu$, Eq.~(\ref{eq:Ps}) immediately leads to  Eq.~(\ref{eq:ps}).

 \end{document}